\documentclass[twocolumn,aps,prb,showpacs,superscriptaddress,floatfix]{revtex4-1}

\usepackage{graphicx}
\usepackage{dcolumn}
\usepackage{bm}
\usepackage{color}
\usepackage{amsmath} 
\usepackage{ae}
\usepackage{float}
\usepackage{footnote}

\newcommand{\angstrom}{\text{\normalfont\AA}}
\newcommand{\bftau}{\mbox{\boldmath$\tau$}}

\begin{document}

\title{Magnetic correlations of the quasi-one-dimensional half-integer spin-chain antiferromagnets Sr$M_2$V$_2$O$_8$ ($M$ = Co, Mn)}

\author{A. K. Bera} 
\email{anup.bera@helmholtz-berlin.de}
\affiliation{Helmholtz-Zentrum Berlin f{\"u}r Materialien und Energie, D-14109 Berlin, Germany}

\author{B. Lake}
\affiliation{Helmholtz-Zentrum Berlin f{\"u}r Materialien und Energie, D-14109 Berlin, Germany}
\affiliation{Institut f{\"u}r Festk{\"o}rperphysik, Technische Universit{\"a}t Berlin, Hardenbergstra{\ss}e 36, D-10623 Berlin, Germany}

\author{W.-D. Stein}
\affiliation{Helmholtz-Zentrum Berlin f{\"u}r Materialien und Energie, D-14109 Berlin, Germany}

\author{S. Zander}
\affiliation{Helmholtz-Zentrum Berlin f{\"u}r Materialien und Energie, D-14109 Berlin, Germany}

\date{\today}

\begin{abstract}
Magnetic correlations of two iso-structural quasi-one-dimensional (1D) antiferromagnetic spin-chain compounds Sr$M_2$V$_2$O$_8$ ({\it M} = Co, Mn) have been investigated by magnetization and powder neutron diffraction. Two different collinear antiferromagnetic (AFM) structures, characterized by the propagation vectors, $k$ = (0 0 1) and $k$ = (0 0 0), have been found below $\sim$ 5.2 K and $\sim$ 42.2 K for the Co-  and Mn-compounds, respectively. For the Mn-compound, AFM chains (along the {\it c} axis) order ferromagnetically within the {\it ab} plane, whereas, for the Co-compound, AFM chains order ferro-/antiferromagnetically along the {\it a/b} direction. The critical exponent study confirms that the Co- and Mn-compounds belong to the Ising and Heisenberg universality classes, respectively. For both compounds, short-range spin-spin correlations are present over a wide temperature range above $T_N$. The reduced ordered moments at base temperature (1.5 K) indicate the presence of quantum fluctuations in both compounds due to the quasi-1D magnetic interactions.

\end{abstract}



\maketitle


\section{Introduction}
Quasi-one dimensional (1D) spin-chain systems are of current interest due to their unconventional magnetic properties.\cite{AffleckJPCM.1.3047} It is well established that an ideal 1D antiferromagnetic (AFM) spin system does not show long-range ordering above $T = 0$ K due to strong quantum spin fluctuations. Moreover, the magnetic properties of such spin-chains are strongly dependent on their spin-values.\cite{HaldanePRL.50.1153,HaldanePLA.93.464} Half-integer spin chains are gapless and a small amount of interchain coupling induces long-range AFM order,\cite{SatijaPRB.21.2001, KojimaPRL.78.1787, DenderPRB.53.2583} while integer spin chains have a singlet ground state and gapped excitations making them stable against magnetic order.\cite{ZheludevPRB.62.8921, BeraPRB.87.224423} In addition, the
presence of anisotropy leads to more complex behavior and a richer phase diagram. 

\begin{figure}
\includegraphics[trim=0cm 0cm 0cm 0cm, clip=true, width=83mm]{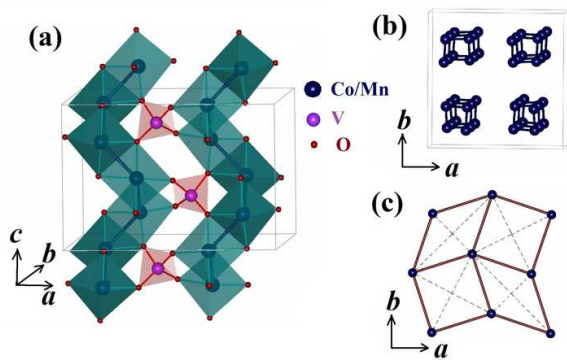}
\caption{\label{Fig:CrysStruc}(Color online) Crystal structure of Sr$M_2$V$_2$O$_8$ ({\it M} = Co, Mn). (a) The propagation of edge shared screw-chains of $M$O$_6$ octahedra along the $c$ axis; for sake of clarity, only two chains (out of four chains per unit cell) are shown. The connections of the chains via VO$_4$ tetrahedra are also shown. The Sr atoms are omitted for simplification. (b) The projection of the four screw-chains per unit cell on to the {\it ab} plane. (c) The geometrical arrangement of the $M^{2+}$ ions in the $ab$ planes. The solid and dashed lines are the $M-M$ connections along the $a/b$ direction and the diagonal directions, respectively.}
 \end{figure}

In this context, the compounds belonging to the material class $AM_2$V$_2$O$_8$ ($A$ = Sr, Ba and $M$ = Cu, Ni, Co, Mn) have recently attracted attention to study the role of anisotropy, spin-value, and interchain interactions on the 1D magnetic properties.\cite{WichmannZAAC.534.153, WichmannRCM.23.1, HePRB.69.220407, BeraPRB.87.224423, HePRB.72.172403, HePRB.73.212406, NiesenJMMM.323.2575} These compounds contain screw-chains of $M$O$_6$ octahedra along the $c$ axis, separated by nonmagnetic VO$_4$ (V$^{5+}$; 3$d^0$, $S = 0$) tetrahedra, resulting in a quasi 1D spin-structure  [Fig. \ref{Fig:CrysStruc}].\cite{BeraPRB.86.024408} The $S$ = 1 AFM chain compound SrNi$_2$V$_2$O$_8$ has a non-magnetic singlet ground state with a gap between the singlet and triplet 1$^{st}$ excited states. The presence of substantial interchain interactions ($zJ_\perp/J \sim$ 0.03; where $z$ is the number of neighbors in other chains) significantly modifies the low energy magnetic excitations and reduces the gap values at the AFM zone centers. \cite{BeraPRB.87.224423,Bera.unpublished} The presence of uniaxial anisotropy (along the $c$ axis) causes a zero field splitting of the triplet states and results in unusual field-dependent magnetic behavior.\cite{BeraPRB.87.224423,Bera.unpublished} Further, the substitution of nonmagnetic ions such as Mg$^{2+}$ at the Ni$^{2+}$ site  induces 3D long-range AFM order.\cite{LappasPRB.66.014428} 

On the other hand, the $S$ = 1/2 XXZ Ising system BaCo$_2$V$_2$O$_8$ undergoes long-range magnetic ordering at 5.4 K in zero magnetic field.\cite{HePRB.72.172403, Kimura.JPCS.51.99, KimuraPRL.99.087602} Application of magnetic field suppresses the AFM order and results in a field-induced order to disorder transition.\cite{HePRB.72.172403} Below 1.5 K, a novel quantum magnetic state with incommensurate propagation vector appears for magnetic fields $\ge$ 4 T. In the present investigation, we have studied the magnetic correlations of two iso-structural quasi-1D half-integer spin-chain compounds Sr$M_2$V$_2$O$_8$  [with {\it M} = Co (effective $S$ = 1/2), Mn ($S$ = 5/2)] to understand the effect of interchain-interactions, anisotropy, and spin-values on the magnetic ground state properties. It is shown that both these compounds develop long-range magnetic order but have different antiferromagnetic ground states. 

Previous reports on bulk magnetic properties showed that the Co- and Mn-compounds order antiferromagnetically below $\sim$ 5 K and $\sim$ 43 K, respectively.\cite{HePRB.73.212406, NiesenJMMM.323.2575} For the Co-compound the magnetic easy axis is the $c$ axis. An interesting field induced order-disorder quantum phase transition was also reported at $\sim$ 3.7 and $\sim$ 6.5 Tesla for $H \parallel c$ and $H \perp c$, respectively, at 2 K which indicates the strongly anisotropic nature of the magnetism.\cite{HePRB.73.212406} It reveals the presence of strong quantum fluctuations which affect the N\'eel AFM state under magnetic field and results in the quantum phase transition. On the other hand,  no such field induced magnetic transition was reported up to 7 Tesla in the Mn-compound. The magnetic ground state of the Mn-compound was predicted to be a classical long-range antiferromagnet with weak anisotropy.\cite{NiesenJMMM.323.2575} The magnetic structures of these compounds, which are necessary to understand their magnetic ground states and microscopic magnetic properties, are currently unknown. 

In this paper, we report a detailed investigation of the magnetic correlations in both the Co- and Mn-compounds by neutron diffraction and dc-magnetization. The magnetic ions Co$^{2+}$ and Mn$^{2+}$  have different electronic configurations (Co$^{2+}$; 3$d^7$, effective $S = 1/2$ and Mn$^{2+}$; 3$d^5$, $S = 5/2$) resulting in different spin values, different ionic radii (0.745 $\angstrom$ for Co$^{2+}$  and 0.83 $\angstrom$ for Mn$^{2+}$)\cite{ShannonACA.32.751} and different anisotropy (Co$^{2+}$ is anisotropic; Ising type and Mn$^{2+}$ is isotropic; Heisenberg type). Antiferromagnetic ordering has been found below $\sim$ 5.2 K and $\sim$ 42.2 K for the Co- and Mn-compounds, respectively. Interestingly, our low temperature neutron diffraction study confirms that they have two completely different collinear AFM structures, characterized by the propagation vectors; $k$ = (0 0 1) and $k$ = (0 0 0) for the Co- and Mn-compounds, respectively. For the Mn-compound, AFM chains (along the $c$ axis) are arranged ferromagnetically within the $ab$ plane, whereas, for the Co-compound, AFM chains are arranged ferro-/antiferromagnetically along the $a/b$ direction.  A critical exponent study was performed to investigate the nature of magnetic ordering in these compounds. Neutron diffraction also reveals the presence of short-range spin-spin correlations over a wide temperature range above $T_N$ for both compounds.

\section{Experimental}
Polycrystalline samples of SrCo$_2$V$_2$O$_8$ were synthesized by using the solid-state reaction method. First the high-purity reagents of SrCO$_3$, CoC$_2$O$_4$.2H$_2$O, and V$_2$O$_5$ are taken as the starting materials in the molar ratio of 1:2:1. The reagents were weighed separately and mixed with ethanol, then ground carefully and homogenized thoroughly in an agate mortar. The mixture was packed into an alumina crucible and calcined at 800~$^\circ$C in air for 72~h with several intermediate grindings. Finally, the product was pressed into pellets and sintered at 900~$^\circ$C in air for 36~h. 

Single crystals of SrCo$_2$V$_2$O$_8$ were grown using the floating zone technique at the crystal laboratory, Helmholtz-Zentrum Berlin f{\"u}r Materialien und Energie (HZB), Berlin, Germany. For this, the powder was pressed into a feed rod using a cold isostatic press. The floating zone furnace was used to melt the tip of the feed rod, which recrystallized on a seed crystal to achieve single-phase growth, the molten zone was then moved along the feed rod at a rate of 1~mm/hour to obtain a large single crystal. The crystals were grown in a mixture of high quality flowing oxygen (20$\%$) and argon (80$\%$) at ambient pressure. The resulting crystals were cylindrical with a diameter of $\sim$ 4~mm and length of $\sim$ 10~cm.

Polycrystalline samples of SrMn$_2$V$_2$O$_8$ were synthesized by a two-step solid-state reaction method. In the first step, as a precursor for the main synthesis, SrV$_2$O$_6$ was synthesized from the mixture of high purity SrCO$_3$ and V$_2$O$_5$ with a 1:1 molar ratio. The chemicals were mixed carefully and homogeneously using a mortar and pestle in ethanol. The mixed chemicals were then put in a Pt-crucible and calcined at 550~$^\circ$C in air for total 24~h with intermediate grindings. For further reaction, the calcined powder was finally heated at 600~$^\circ$C for 48~h. In the second step, the as-prepared SrV$_2$O$_6$ and high purity MnO were mixed in a 1:2 molar ratio and the mixture was ground homogeneously in an agate mortar. The mixture was then pressed into pellets and sintered at 850~$^\circ$C in Ar flow for total of 144~h with intermediate grindings and repelletizations. 

The as-prepared powder samples were characterized by powder X-ray diffraction (XRD) using a lab X-ray machine (Bruker D8 Advance). The XRD measurements were performed at 300~K using Cu $K_\alpha$ radiation over the scattering angular range 2$\theta =12^\circ$--$ 90^\circ$. The phase purity of the single crystals were also confirmed by powder XRD. A part of the crystal was ground into powder for this study. The backscattering X-ray Laue patterns revealed the good single-crystalline nature of the samples.

Temperature and field-dependent magnetization measurements were carried out using a Physical Properties Measurement System (14~T PPMS; Quantum Design) at the Laboratory for Magnetic Measurements, HZB. The temperature-dependent static susceptibility [$\chi$(T)] measurements were performed over a temperature range of 2--400~K for the Co-compound and 2--900~K for the Mn-compound, respectively, under an applied magnetic field of 1~Tesla. All measurements were performed in the warming cycle after cooling the samples in zero field. The isothermal magnetization measurements were performed at the base temperature 1.8~K over $\pm 14$~Tesla in the zero field cooled condition. Single crystal measurements were carried out on a small piece of the crystal (3 $\times$ 2 $\times$ 1 mm$^3$) with edges parallel to the principal crystallographic axes.

For the crystal structure investigation, neutron diffraction patterns for both compounds were recorded at 300~K by using the high resolution powder diffractometer E9 ($\lambda$ = 1.7982~$\angstrom$), at HZB, Germany. For the magnetic structure investigation, low temperature neutron diffraction patterns were measured by using the time-of-flight diffractometer V15, at HZB, Germany. V15 is equipped with four movable detector banks which were positioned at 30, 60, 90, and 150 degrees. The incident wavelength band 0.7--7.8~$\angstrom$ was used for the measurements. Low temperature measurements were performed in a standard orange cryostat. The diffraction data were analyzed by the Rietveld method using the FULLPROF program.\cite{Fullprof}

\section{results and discussion}
\subsection{Crystal structure}

\begin{figure}
\includegraphics[trim=4.0cm 6.5cm 6.3cm 0cm, clip=true, width=80mm]{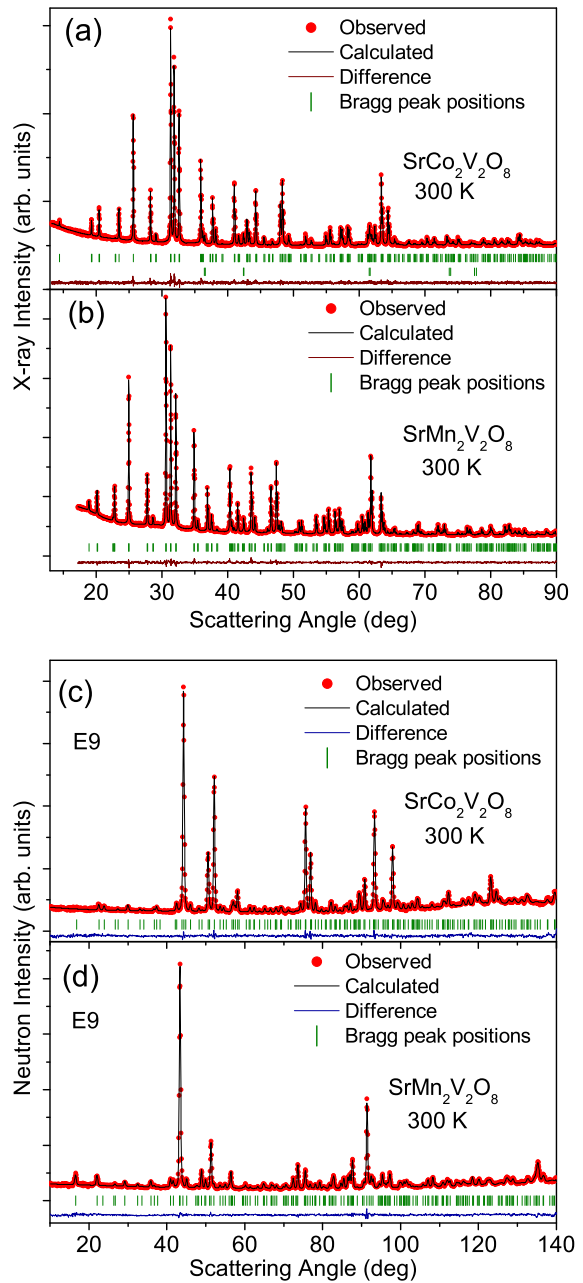}
\caption{\label{Fig:xrd}(Color online) The observed (filled circles) and calculated (solid black line) X-ray [(a) and (b)] and neutron [(c) and (d)] diffraction patterns for SrCo$_2$V$_2$O$_8$ and SrMn$_2$V$_2$O$_8$. The difference between observed and calculated patterns are shown by thin lines at the bottom of each panel. The vertical bars are the allowed Bragg peak positions. The second row of the peak markers in (a) are for the CoO impurity phase.}
\end{figure}

\begin{table}
\caption{\label{tab:T1}The Rietveld refined lattice constants ($a$ and $c$), fractional atomic coordinates, and isotropic thermal parameters ($B_{iso}$) for Sr$M$$_2$V$_2$O$_8$ ($M$ = Co, Mn) at 300~K.}
\begin{ruledtabular}
\begin{tabular}{cccc}
 &Site&SrCo$_2$V$_2$O$_8$&SrMn$_2$V$_2$O$_8$\\
\hline
$a ($\angstrom$)$ &  &12.2710(1) & 12.4527(1)\\
$c ($\angstrom$)$ &  &8.4192(1) & 8.6853(1)\\
\\
Sr & 8$a$ (0,0,$z$)& & \\
$z/c$& & 0 & 0\\
$B_{iso}$& & 1.41(12) & 0.95(11)\\
\\
$M$ (Co/Mn) & 16$b$ ($x$,$y$,$z$)& & \\
$x/a$& & 0.3281(12) & 0.3282(8)\\
$y/b$& & 0.3311(9) & 0.3308(6)\\
$z/c$& & 0.2147(18) & 0.2107(11)\\
$B_{iso}$& & 0.62(12) & 0.66(9)\\
\\
V & 16$b$  ($x$,$y$,$z$)& & \\
$x/a$& & 0.2615(3) & 0.2638(9) \\
$y/b$& & 0.0783(4) & 0.0786(4)\\
$z/c$& & 0.0942(9) &  0.0776(3)\\
$B_{iso}$& & 0.31(9) & 0.27(2)\\
\\
O1 & 16$b$  ($x$,$y$,$z$)& & \\
$x/a$& & 0.1457(4) & 0.1399(4) \\
$y/b$& &0.5014(7) & 0.4990(7)\\
$z/c$& & -0.0026(12) &  -0.0087(10)\\
$B_{iso}$& & 1.26(12) & 1.22(11)\\
\\
O2 & 16$b$  ($x$,$y$,$z$)& & \\
$x/a$& & 0.3410(6) & 0.3419(5) \\
$y/b$& & 0.6630(7) &  0.6668(6)\\
$z/c$& & 0.4763(9) &  0.4622(8)\\
$B_{iso}$& & 0.49(16) & 0.61(12)\\  
\\
O3 & 16$b$  ($x$,$y$,$z$)& & \\
$x/a$& & 0.1578(8) & 0.1499(6) \\
$y/b$& & 0.6824(5) &  0.6885(4)\\
$z/c$& & 0.7086(9) &  0.7009(8)\\
$B_{iso}$& & 0.58(17) & 0.90(11) \\ 
\\
O4 & 16$b$  ($x$,$y$,$z$)& & \\
$x/a$& & 0.3314(5) & 0.3266(4) \\
$y/b$& &  0.4974(6) &  0.4996(6)\\
$z/c$& & 0.1876(9) &  0.1727(7)\\
$B_{iso}$& &  1.02(12)  & 0.80(10) \\
\end{tabular}
\end{ruledtabular}
\end{table}

The crystal structures of SrCo$_2$V$_2$O$_8$ and SrMn$_2$V$_2$O$_8$ are investigated from simultaneous Rietveld analysis of both X-ray and neutron powder diffraction patterns, recorded at 300~K (Fig. \ref{Fig:xrd}). Rietveld analysis confirms that both compounds crystallize in the tetragonal space group $I4_1cd$. It also confirms that the Mn-compound is single phase in nature. The Co-compound used for the X-ray and low temperature neutron diffraction measurements contains a small amount ($\sim$ 4 $\%$) of CoO impurity phase. However, a phase pure sample of SrCo$_2$V$_2$O$_8$ (without CoO impurity) was used for the neutron diffraction study at 300~K. The refined values of various structural parameters such as lattice constants, fractional atomic coordinates, and isotropic thermal parameters are given in Table \ref{tab:T1}. 

The values of the lattice constants are found to be $a = 12.2710(1)$~$\angstrom$ and $c = 8.4192(1)$~$\angstrom$ for the Co-compound and $a = 12.4527(1)$~$\angstrom$ and $c = 8.6853(1)$~$\angstrom$ for the Mn-compound, respectively. These values are in good agreement with the values reported earlier.\cite{HeJCG.293.458,LejayJCG.317.128,NiesenJMMM.323.2575} The larger lattice constants of the Mn-compound compared to the Co-compound are due to their ionic radii. The greater ionic radius of the Mn$^{2+}$ ion (0.83~$\angstrom$) compared to that of the Co$^{2+}$ ion (0.745~$\angstrom$) in octahedral coordination causes this increase of the lattice constants. The isostructural compound SrNi$_2$V$_2$O$_8$ also follows this sequence. Its lattice constants were reported to be $a = 12.1608(1)$~$\angstrom$ and $c = 8.3242(1)$~$\angstrom$ which are the smallest in this series due to the smaller ionic radius of the Ni$^{2+}$ ions (0.69~$\angstrom$).\cite{BeraPRB.86.024408}

\begin{table}
\caption{\label{tab:T2}The bond lengths and $M$-$M$ distances in Sr$M$$_2$V$_2$O$_8$ ($M$ = Co, Mn) at 300~K.}
\begin{ruledtabular}
\begin{tabular}{ccc}
Bonds & SrCo$_2$V$_2$O$_8$ & SrMn$_2$V$_2$O$_8$\\
\hline
Sr-O1 ($\angstrom$) & 2$\times$2.779(10) & 2$\times$2.725(9)\\
                              & 2$\times$2.746(9)  &  2$\times$2.843(9)\\
Sr-O2 ($\angstrom$) & 2$\times$2.801(8) & 2$\times$2.881(6)\\
Sr-O3 ($\angstrom$) & 2$\times$2.980(8) & 2$\times$3.029(7)\\
Sr-O4 ($\angstrom$) & 2$\times$2.603(7) & 2$\times$2.629(5)\\
                              & 2$\times$3.346(7) & 2$\times$3.571(6)\\
\\
$M$-O1 ($\angstrom$) & 2.162(16) & 2.176(11)\\
$M$-O2 ($\angstrom$) & 2.015(17) & 2.165(12)\\
                                 & 2.119(14) & 2.152(10)\\
$M$-O3 ($\angstrom$) & 2.097(18) & 2.234(12)\\
                                 & 2.062(17) & 2.110(12)\\
$M$-O4 ($\angstrom$) & 2.051(13) & 2.128(11)\\
\\
V-O1 ($\angstrom$) & 1.689(8) & 1.727(7)\\
V-O2 ($\angstrom$) & 1.753(8) & 1.758(7)\\
V-O3 ($\angstrom$) & 1.788(8) & 1.697(7)\\
V-O4 ($\angstrom$) & 1.704(7) & 1.708(6)\\
\\
$M$-$M$ ($\angstrom$) (along $c$ axis) & 2.874(19) & 2.937(14)\\
$M$-$M$ ($\angstrom$) (along $a/b$ axis) & 6.428(15) & 6.523(11)\\
                                                             & 6.452(13) & 6.543(12)\\
\end{tabular}
\end{ruledtabular}
\end{table}
                         
The crystal structure of these compounds consists of screw chains along the $c$ axis [Fig. \ref{Fig:CrysStruc}]. The screw chains are formed by corner sharing $M$O$_6$ octahedra and are connected via VO$_4$ tetrahedra by sharing corners  [Fig. \ref{Fig:CrysStruc}(a)]. The $M$O$_6$ octahedra and VO$_4$ tetrahedra are found to be slightly distorted as evident from the differences in the Co/Mn-O and V-O bond lengths [Table \ref{tab:T2}]. There are four screw chains per unit cell which are centered around (1/4,1/4), (1/4,3/4), (3/4,1/4), and (3/4,3/4) in the $ab$ plane [Fig. \ref{Fig:CrysStruc}(b)]. Two diagonal chains rotate clockwise while  the other two chains rotate anti-clockwise when propagating along the $c$ axis which results in a complex interchain interaction geometry within a given $ab$ plane [Fig. \ref{Fig:CrysStruc}(c)]. Furthermore the interaction geometry rotates 90$^\circ$ in the next layer (shifted by $c$/4) along the $c$ axis. Interchain interactions are also possible, via superexchange pathways (through the VO$_4$ tetrahedra), from Mn/Co ions in a chain to the Mn/Co ions which are shifted by $c$/4 in the neighboring chains [Fig. \ref{Fig:CrysStruc}]. The peculiar crystal structure, therefore, provides many possible interchain super-exchange interaction pathways with comparable strengths. The distances between magnetic ions along the chain direction are 2.874 $\pm$ 0.019~$\angstrom$ for the Co-compound and 2.937 $\pm$ 0.014~$\angstrom$ for the Mn-compound, respectively. Within a given $ab$ (along $a/b$ axis) plane, the distances between magnetic ions are 6.428 $\pm$ 0.015/6.452 $\pm$ 0.013~$\angstrom$ for the Co-compound and 6.523 $\pm$ 0.011/6.543 $\pm$ 0.012~$\angstrom$ for the Mn-compound, respectively. The different interchain distances for the two compounds are also due to the ionic size effect.

\subsection{Magnetization}

\begin{figure}
\includegraphics[trim=3.0cm 4cm 13.0cm 1.5cm, clip=true, width=80mm]{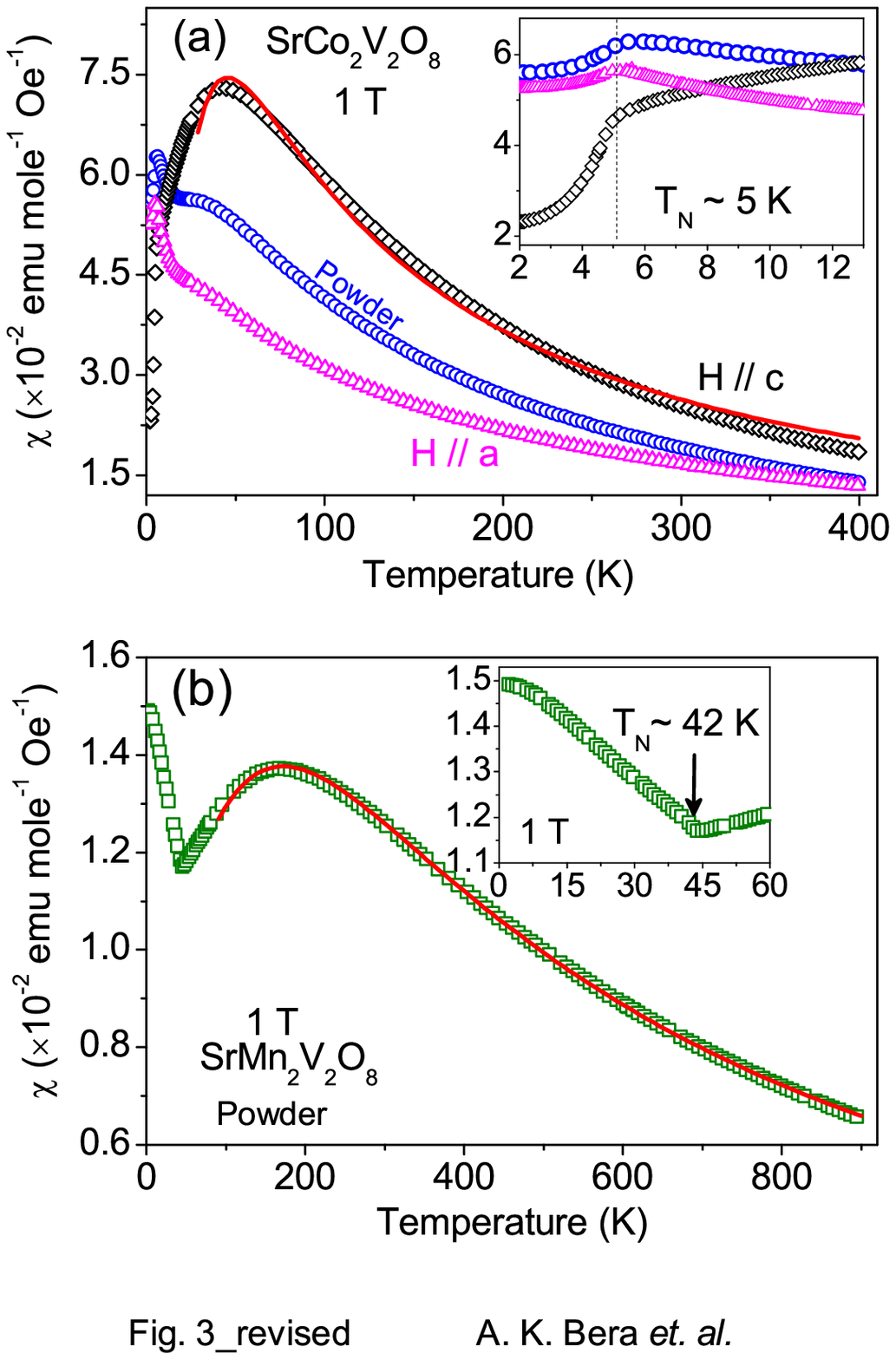}
\caption{\label{Fig:Sus}(Color online) The temperature-dependent susceptibility ($\chi$) curves (open symbols) for (a) SrCo$_2$V$_2$O$_8$ and (b) SrMn$_2$V$_2$O$_8$ compounds, respectively, measured under 1~Tesla of magnetic field. The solid lines in each panels are fitted curves as per the models (Eq. \ref{eq:ising-susc-para} for the Co- and  Eq. \ref{eq:pade} for the Mn-compounds, respectively. The insets of both panels show the enlarged view of the $\chi$($T$) curves around the transition temperatures. }
 \end{figure}

Figures \ref{Fig:Sus}(a) and \ref{Fig:Sus}(b) show the susceptibility ($\chi$) vs. temperature curves, measured under a magnetic field of 1~T, for  SrCo$_2$V$_2$O$_8$ and SrMn$_2$V$_2$O$_8$, respectively. For  SrCo$_2$V$_2$O$_8$, with decreasing temperature a broad hump with maximum around 30~K followed by a sharp peak at $\sim$ 5.2~K [Fig. \ref{Fig:Sus}(a)] has been found for the powder sample. The single crystal susceptibilities show that the broad peak $\sim$ 30~K is present for both parallel ($H$ $\parallel$ $c$) and perpendicular ($H$ $\parallel$ $a$) field directions. With decreasing temperature, below $\sim$ 5.2~K, the parallel susceptibility ($H$ $\parallel$ $c$) shows a sudden decrease, whereas, a sharp peak appears in the perpendicular susceptibility ($H$ $\parallel$ $a$) at  $\sim$ 5.2~K [Fig. \ref{Fig:Sus}(a)]. The broad hump is characteristic of the onset of 1D short-range spin-spin correlations and the sharp peak corresponds to the onset of 3D long-range AFM ordering. The presence of short-range spin-spin correlations is confirmed from the neutron diffraction study (see section E). A large difference between parallel and perpendicular susceptibilities has been observed that persists up to 400~K. This indicates the presence of a large paramagnetic anisotropy. 

It is also found that the $\chi(T)$ curves do not follow Curie-Weiss behavior below 400~K which may be due to low-lying crystal field excitations as well as the presence of short-range correlations. It should be noted that the $\chi(T)$ curves do not show a rapid jump $\sim$ 20~K [Fig. \ref{Fig:Sus}(a)] which was reported by He {\it et. al}.,\cite{HePRB.73.212406} and was explained on the basis of weak ferromagnetic moments due to canted antiferromagnetic ordering. In agreement with our results, the susceptibility curves reported by Lejay {\it et. al}.\cite{LejayJCG.317.128} from a single crystal study do not show this jump.  In addition, no ferromagnetism is evident in our low temperature neutron diffraction results (discussed in section C).

It is known that for the free Co$^{2+}$ (3$d^7$) ion having 7 electrons the total orbital and spin angular momenta are $L$ = 3 and $S$ = 3/2, respectively, according to Hund's rules. A cubic crystal field splits the 7 orbital levels into 1 orbital singlet and 2 orbital triplets. The lowest state is a triplet $T_1$.  The effective orbital angular momentum within the triplet $T_1$ is $L$ = 1. In octahedral ligand field and with spin-orbit coupling the degeneracy of the 12 states in $T_1$ is partially lifted and results in 6 Kramers doublets. The lowest Kramers doublet can be regarded as an effective $S$ = 1/2 with large anisotropy. The exchange interaction between true Co$^{2+}$ spins ($S$ = 3/2) may be regarded as a molecular field which lifts the degeneracy of the Kramers doublets. 

To estimate the intrachain interaction in SrCo$_2$V$_2$O$_8$, the parallel susceptibility curve is fitted with the Bonner-Fisher model for uncoupled Ising chains (Ref. \onlinecite{BonnerPR.135.A640}) 

\begin{eqnarray}
\chi_\parallel (T) = \frac{N_A \mu_B^2 g_\parallel^2}{4k_B T}exp(\frac{J_{NN}}{k_B T})
\label{eq:ising-susc-para}
\end{eqnarray}

\noindent where $g_\parallel$ is the Land{\'e} factor parallel to the Ising axis, $k_B$ stands for the Boltzmann constant and $J_{NN}$ is the nearest neighbor intrachain exchange constant. The fitted values of the parameters are found to be   $J_{NN}$ = --45.35 $\pm$ 0.04~K and $g_\parallel$ = 9.90 $\pm$ 0.01. The values of $J_{NN}$ = --40.6~K and $g_\parallel$ = 9.42 were reported by He {\it et. al}.\cite{HePRB.73.212406} The fitted curve is shown by the solid line. The discrepancy between the observed and fitted curves indicates that the present compound deviates from the pure Ising limit and may be better defined as an XXZ-system. 

For isolated chains, the perturbation from the pure Ising limit gives rise to a continuum of soliton-pair excited states. This is in contrast to the doubly degenerate ground states and highly degenerate set of first excited states of energy 2$J$ in the pure Ising limit.\cite{GoffPRB.52.15992} The continuum states may be split into discrete levels by interactions with adjacent chains as reported for CsCoCl$_3$.\cite{GoffPRB.52.15992} The mixing of crystal-field levels of the Co$^{2+}$ ion with the spin excitation states may also be expected which leads to a decrease of the energy values of the excited states.

For SrMn$_2$V$_2$O$_8$, with decreasing temperature the susceptibility curve shows a broad maximum $\sim$ 170~K, characteristic of short-range 1D magnetism, followed by a kink $\sim$ 42~K corresponding to the 3D long-range AFM transition [Fig. \ref{Fig:Sus}(b)]. The nature of the susceptibility curve is similar to that reported earlier by Niesen {\it et. al.} for this compound.\cite{NiesenJMMM.323.2575} Below T$_N$, the susceptibility curve showed a slight increase, which was ascribed to the transverse susceptibility.\cite{NiesenJMMM.323.2575} To estimate the strength of the intrachain interactions, the susceptibility curve above 90~K is fitted with a general expression of the susceptibility of Heisenberg spin-chains predicted from Pad\'e approximations \cite{LawJPCM.25.065601}

\begin{eqnarray}
\chi_{obs}(T) = \chi_0 +\frac{N_A \mu_B^2 g^2 S(S+1)}{3k_B T}\nonumber  \\
\times\frac{1+\sum_{i=1}^m A_i (\frac{J_{NN}}{k_B T})^i}
{1+\sum_{j=1}^n B_j (\frac{J_{NN}}{k_B T})^j} 
\label{eq:pade}
\end{eqnarray} 

\noindent where the first temperature-independent term $\chi_0$ is due to Van-Vleck paramagnetism as well as diamagnetic core susceptibility and $J_{NN}$ is the nearest-neighbor intrachain super-exchange interaction. A good agreement between observed and calculated susceptibility curves is found. The fitted values of the parameters are $J_{NN}$ = 37.57 $\pm$ 0.02~K, $g$ = 1.9999 $\pm$ 0.0009, and $\chi_0$ = $-3.67\times10^{-4} \pm 0.07\times10^{-4}$ emu mole$^{-1}$ Oe $^{-1}$.  For the fitting, the values of the coefficients $A_i$ and $B_j$ are adopted from Ref. \onlinecite{LawJPCM.25.065601}. 

\begin{figure}
\includegraphics[trim=2.5cm 4cm 13.5cm 1.5cm, clip=true, width=80mm]{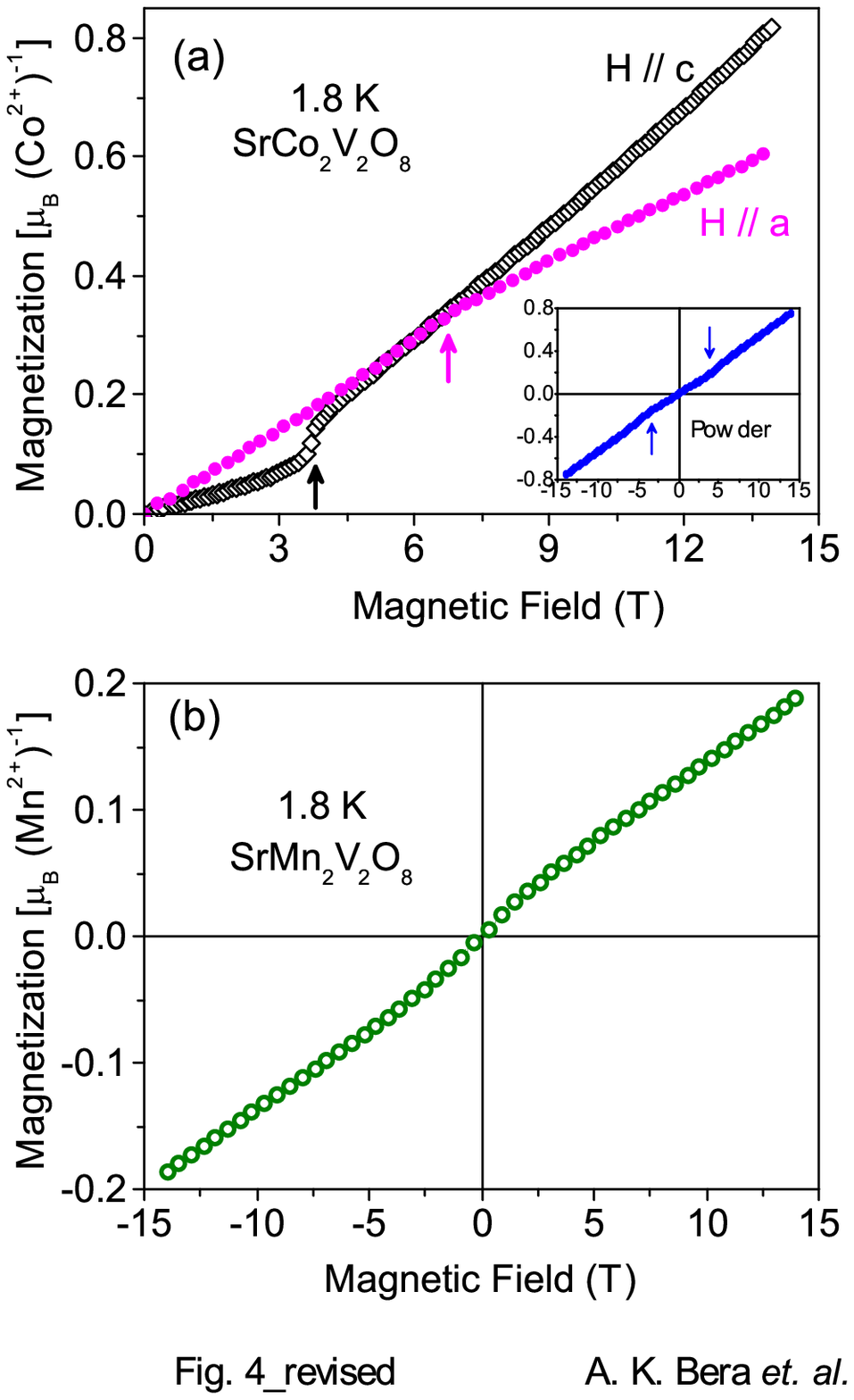}
\caption{\label{Fig:Mag}(Color online) The magnetization vs. magnetic field curves for (a) SrCo$_2$V$_2$O$_8$ and (b) SrMn$_2$V$_2$O$_8$ compounds, respectively, measured at 1.8~K. The arrows in the upper panel show the field induced transitions. The inset shows the M vs H curves for the powder sample. The arrows indicate the onset of the field induced transitions.}
 \end{figure}

The field dependence of the magnetizations for both compounds, measured at 1.8~K are shown in Fig. \ref{Fig:Mag}. For SrCo$_2$V$_2$O$_8$, field induced phase transitions under the critical fields of $\sim$ 3.7 and $\sim$ 6.5~T for $H \parallel c$ and $H \perp c$ ($H \parallel a$), respectively, have been found. Same critical field values ($H_c^\parallel$ = 3.7~T and $H_c^\perp$ = 6.5~T) were reported earlier by He {\it et. al.}\cite{HePRB.73.212406} For the powder sample, as expected, a broad transition extending from 3.7~T to 6.5~T is found for both positive and negative magnetic fields [inset of Fig. \ref{Fig:Mag}(a)]. 

A similar field induced magnetic (order-disorder) transition was reported in the related compound BaCo$_2$V$_2$O$_8$ for $H \parallel c$ and was attributed to the quantum phase transition characteristic of the $S$ = 1/2 1D XXZ antiferromagnet with Ising-like anisotropy. \cite{Kimura.JPCS.51.99,KimuraPRL.99.087602} Here, the magnetic field induces a reentrant phenomenon from the 3D long-range N\'eel state into a 1D quantum spin-liquid state. This transition was theoretically predicted first for gapped spin systems as a function of doping of the magnetic site by non-magnetic impurities. \cite{MikeskaPRL.93.217204}  The magnetization of SrCo$_2$V$_2$O$_8$ shows an almost linear field dependence below the transition fields which indicates the absence of a ferromagnetic contribution in agreement with the neutron diffraction results (discussed later). No hysteresis and remanent magnetization were found at 1.8~K. There is also no tendency towards saturation up to 14~T. 

On the other hand, for SrMn$_2$V$_2$O$_8$ no such field induced transition has been found. In this case, the magnetization shows almost linear behavior with magnetic field. The value of magnetization is small $\sim 0.2~\mu_B$/Mn$^{2+}$ at 14~T as compared to the fully polarized magnetization of 5~$\mu_B$/Mn$^{2+}$ which confirms the presence of strong AFM interactions. This is in agreement with neutron diffraction results where an AFM ordering has been found for this compound (discussed later). The magnetization studies clearly reveal that two compounds have very different magnetic properties despite their similar crystal structures.

\subsection{Magnetic structures}

In order to investigate the nature of the ground state for these two iso-structural compounds, we have carried out low temperature neutron diffraction measurements at several temperatures below and above the corresponding N\'eel temperatures. For this purpose, we employed the time-of-flight diffractometer V15 which provides good resolution over the low $Q$ (high $d$) region. 

\subsubsection{Magnetic structure of SrCo$_2$V$_2$O$_8$}

\begin{figure}
 \includegraphics[trim=4.2cm 4.7cm 13.5cm 0.5cm, clip=true, width=80mm]{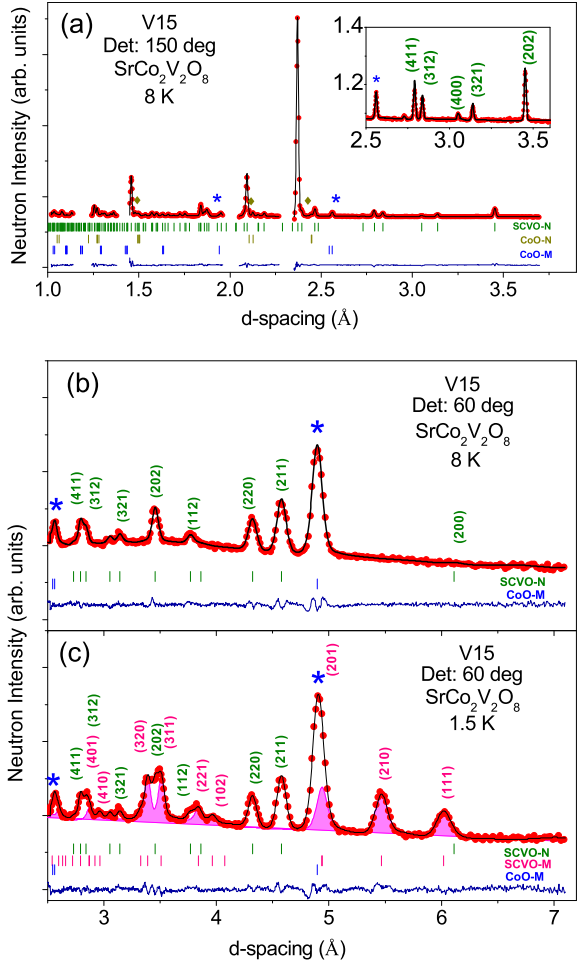}
 \caption{\label{Fig:Neu-SCVO}(Color online) The neutron diffraction patterns of SrCo$_2$V$_2$O$_8$ at (a) 8~K (from the detector bank centered at 150~deg),  (b) 8~K (from the detector bank at 60~deg), and (c) 1.5~K (from the detector bank at 60~deg). The observed and calculated patterns are shown by filled circles and solid black lines, respectively. The difference patterns between observed and calculated patterns are shown by solid lines at the bottom of each panel. The vertical bars are the calculated positions of the Bragg peaks. Different colors of the vertical bars correspond to the different phases. For the refinement of the diffraction pattern from the detector bank at 150~deg, shown in (a), few regions are excluded where the  aluminum lines (from the cryostat) were observed. The nuclear and magnetic Bragg peaks from CoO are marked with diamonds and stars, respectively. The inset of (a) shows an enlarged view of the diffraction pattern over the high {\it d}-value region. The shaded regions in (c) show the contributions of the magnetic scattering from SrCo$_2$V$_2$O$_8$.}
 \end{figure}

The neutron diffraction patterns of SrCo$_2$V$_2$O$_8$ at 8~K (paramagnetic state) and 1.5~K  (magnetically ordered state) are shown in Fig. \ref{Fig:Neu-SCVO} [(a)--(b)] and Fig. \ref{Fig:Neu-SCVO}[(c)], respectively. Owing to the presence of a small CoO impurity in this sample (discussed earlier), the neutron diffraction patterns at 8~K are refined with a model which included three phases. The phases are (i) nuclear phase for SrCo$_2$V$_2$O$_8$ and (ii) nuclear as well as  (iii) magnetic phases for CoO. The CoO orders antiferromagnetically below 295~K.\cite{JauchPRB.64.052102} First we refined the diffraction pattern from the detector bank at 150~deg [Fig. \ref{Fig:Neu-SCVO}(a)]  which covers the $d$-range from 1--3.7~$\angstrom$. Good agreement between observed and calculated patterns has been obtained [Fig. \ref{Fig:Neu-SCVO}(a)]. The tetragonal crystal structure with space group $I4_1cd$ (as found at room temperature) reproduces the low temperature nuclear phase of SrCo$_2$V$_2$O$_8$. The refined values of the lattice parameter are found to be $a = 12.206 \pm 0.011$~$\angstrom$  and $c = 8.374 \pm  0.013$~\angstrom, respectively. 

The crystal structure for CoO is monoclinic with space group $C2/m$.\cite{JauchPRB.64.052102} The lattice parameters are found to be $a$ = 5.168 $\pm$ 0.018~\angstrom, $b$ = 3.004 $\pm$ 0.043~\angstrom, $c$ = 3.008 $\pm$ 0.023~\angstrom, and $\beta$ = 125.5 $\pm$ 0.8~deg which are in good agreement with the reported values. \cite{JauchPRB.64.052102} The AFM structure with propagation vector {\it k} = (0 1 1/2) was used for the magnetic phase of CoO as reported by Jauch {\it et. al.} \cite{JauchPRB.64.052102} The refined value of the ordered magnetic moment is 3.9 $\pm$ 0.1~$\mu_B$/Co$^{2+}$ which also agrees well with the reported value of 3.98~$\mu_B$/Co$^{2+}$ at 10~K.\cite{JauchPRB.64.052102} The obtained values of structural parameters and magnetic moment were used to refine the diffraction patterns from the detector bank at 60~deg which covers high {\it d}-value region (2.5--7.2~\angstrom) and  important for the magnetic scattering study of SrCo$_2$V$_2$O$_8$. The refined diffraction pattern from this detector bank at 8~K is depicted in Fig. \ref{Fig:Neu-SCVO}(b) and shows good agreement between observed and calculated patterns.
        
The diffraction pattern at the base temperature of 1.5~K (magnetic ordered state) is shown in Fig. \ref{Fig:Neu-SCVO}(c). Appearance of a set of additional magnetic Bragg peaks has been observed which are at positions ($hkl$) with $h+k+l$ is an odd integer, that is, on the nuclear peaks forbidden by the body-centered symmetry $I$ of the lattice. All magnetic peaks can be indexed with a propagation vector $k$ = (0 0 1) with respect to the tetragonal symmetry. 

In order to find the magnetic structure that is compatible with the space group symmetry, we performed representation analysis by using version 2K of the program SARAh-Representational Analysis.\cite{WillsphysicaB.276.680} The analysis was done by using the crystal structure above the transition and the propagation vector of the magnetic ordering. Firstly it involves the determination of the space group symmetry elements, that leave the propagation vector $k$ invariant: these form the little group $G_k$. The magnetic representation of a crystallographic site can then be decomposed in terms of the irreducible representations (IRs) of $G_k$: 

\begin{equation}
 \Gamma_{mag}=\sum_{\nu} n_{\nu}\Gamma_{\nu}^{\mu}
 \label{magnetic_representation}
\end{equation}

\noindent where $n_\nu$ is the number of times that the IR $\Gamma_\nu$ of order $\mu$ appears in the magnetic representation $\Gamma_{mag}$ for the chosen crystallographic site.

\begin{table}
\caption{\label{tab:T3}Irreducible representations of the group  $G_k$ of the propagation vector $k$ = (0 0 1) for SrCo$_2$V$_2$O$_8$. The atoms of the nonprimitive basis are defined according to Co1: (0.3315, 0.3301, 0.2233), Co2: (0.6685, 0.6699, 0.2233), Co3: (0.6699, 0.8315, 0.4733), Co4: (0.3301, 0.1685, 0.4733), Co5: (0.6685, 0.3301, 0.7233), Co6: ( 0.3315, 0.6699, 0.7233), Co7: (0.8301, 0.3315, 0.4733), Co8: (0.1699, 0.6685, 0.4733).}
\begin{ruledtabular}
\begin{tabular}{ccccccccc}
IRs & & & & Atoms & & & & \\
& Co1 & Co2 & Co3 & Co4 & Co5 & Co6 & Co7 & Co8 \\
\hline
$\Gamma_1^1$ & 1 & 1 & -i & -i & -1 & -1 & -i & -i \\ \\
$\Gamma_2^1$ & 1 & 1 & -i & -i &  1 &  1 &  i &  i \\ \\
$\Gamma_3^1$ & 1 & 1 &  i &  i & -1 & -1 &  i &  i \\ \\
$\Gamma_4^1$ & 1 & 1 &  i &  i &  1 &  1 & -i & -i \\ \\
$\Gamma_5^2$ & 1  0 & -1  0 & 1  0 & -1  0 & 0  -1 & 0  1 & 0  -1 & 0  1 \\
 & 0 1 & 0  -1 & 0  -1 & 0  1 & -1  0 &  1  0 &  1  0 & -1  0 \\
\end{tabular}
\end{ruledtabular}
\end{table}

 For SrCo$_2$V$_2$O$_8$, the crystal structure is described by the space group $I4_1 c d$ (Sp. Gr. No. 110). This space group involves two centering operations and eight symmetry operations (Appendix \ref{G0_elements}). All the eight symmetry operations leave the propagation vector $k$ invariant or transform it into an equivalent vector. For the propagation vector $k$ = (0 0 1), the irreducible representations of the propagation vector group $G_k$ are given in Table \ref{tab:T3}. There are five possible IRs. The magnetic reducible representation $\Gamma_{mag}$ for the Co site (16$b$ site) can be decomposed as a direct sum of IRs as
\begin{equation}
\Gamma_{mag}=3\Gamma_{1}^{1}+3\Gamma_{2}^{1}+3\Gamma_{3}^{1}+3\Gamma_{4}^{1}+6\Gamma_{5}^{2}
\label{magnetic_representation_SCVO}
\end{equation}

\begin{table}
\caption{\label{tab:T4}Basis Vectors of position 16$b$ for the IRs with $k$ = (0 0 1) for SrCo$_2$V$_2$O$_8$. Only the real components of the basis vectors are presented. The atomic sites are labeled following the convention given in Table \ref{tab:T3}.}
\begin{ruledtabular}
\begin{tabular}{cccccccccc}
IRs &&\multicolumn{8}{c}{Basis Vectors} \\
& &Co1 & Co2 & Co3 & Co4 & Co5 & Co6 & Co7 & Co8 \\
\hline
$\Gamma_1^1$ & $\Psi_1$ &(200)&(-200)&(000)&(000)&(-200)	&(200)&(000)&	(000)\\
 & $\Psi_2$ &(020)&(0-20)&(000)&(000)&(020)&(0-20)&(000)&(000)\\
 & $\Psi_3$ &(002)&(002)&(000)&(000)&(002)&(002)&(000)&(000)\\
\\
$\Gamma_2^1$ &$\Psi_1$ &(200)&	(-200)&	(000)&	(000)&	(200)&	(-200)&	(000)&	(000) \\
&$\Psi_2$&(020)&	(0-20)&	(000)&	(000)&	(0-20)&	(020)&	(000)&	(000) \\
&$\Psi_3$&(002)&	(002)&	(000)&	(000)&	(00-2)&	(00-2)&	(000)&	(000) \\
\\
$\Gamma_3^1$ & $\Psi_1$ & (000)&	(000)&	(020)&	(0-20)&	(000)&	(000)&	(0-20)&	(020) \\
&$\Psi_2$&(000)&	(000)&	(-200)&	(200)&	(000)&	(000)&	(-200)&	(200) \\
&$\Psi_3$&(000)&	(000)&	(002)&	(002)&	(000)&	(000)&	(00-2)&	(00-2) \\
\\
$\Gamma_4^1$ &  $\Psi_1$  & (000)	&(000)&	(020)&	(0-20)&	(000)&	(000)&	(020)&	(0-20) \\
& $\Psi_2$&(000)&	(000)	&(-200)&	(200)&	(000)&	(000)&	(200)&	(-200) \\
& $\Psi_3$&(000)&	(000)&	(002)&	(002)&	(000)&	(000)&	(002)&	(002) \\
\\
$\Gamma_5^2$ &$\Psi_1$& (100)&	(100)&	(010)&	(010)&	(000)&	(000)&	(000)&	(000) \\
&$\Psi_2$&(010)&	(010)&	(-100)&	(-100)&	(000)&	(000)&	(000)&	(000) \\
&$\Psi_3$&(001)&	(00-1)&	(001)&	(00-1)&	(000)&	(000)&	(000)&	(000) \\
&$\Psi_4$&(000)&	(000)&	(000)&	(000)&	(-100)&	(-100)&	(010)&	(010) \\
&$\Psi_5$&(000)&	(000)&	(000)&	(000)&	(010)&	(010)&	(100)&	(100) \\
&$\Psi_6$&(000)&	(000)&	(000)&	(000)&	(001)&	(00-1)&	(001)&	(00-1) \\
&$\Psi_7$&(000)&	(000)&	(000)&	(000)&	(-100)&	(-100)&	(0-10)&	(0-10) \\
&$\Psi_8$&(000)&	(000)&	(000)&	(000)&	(010)&	(010)&	(-100)&	(-100) \\
&$\Psi_9$&(000)&	(000)&	(000)&	(000)&	(001)&	(00-1)&	(00-1)&	(001) \\
&$\Psi_{10}$&(100)&	(100)&	(0-10)&	(0-10)&	(000)&	(000)&	(000)&	(000) \\
&$\Psi_{11}$&(010)&	(010)&	(100)&	(100)&	(000)&	(000)&	(000)&	(000) \\
&$\Psi_{12}$&(001)&	(00-1)&	(00-1)&	(001)&	(000)&	(000)&	(000)&	(000) \\
\end{tabular}
\end{ruledtabular}
\end{table}

According to the Landau theory for second order phase transitions, only one representation can be involved in a critical transition. With this constraint, therefore, five different magnetic structures are possible with $k$ = (0 0 1). The basis vectors of the Co position [16$b$ ($x$, $y$, $z$); (0.3315, 0.3301, 0.2233)] for five IRs, calculated using the projection operator technique, implemented in SARAh,\cite{WillsphysicaB.276.680} are given in Table \ref{tab:T4}. The labeling of the propagation vector and the IRs follows the scheme used by Kovalev.\cite{Kovalev.book} The atomic sites are labeled following the convention given in Table \ref{tab:T3}. 

The representations $\Gamma_{1}$, $\Gamma_{2}$, $\Gamma_{3}$, and $\Gamma_{4}$ are one dimensional and are repeated three times each in Eq. \ref{magnetic_representation_SCVO}. There are three basis vectors corresponding to each of these representations. Moreover, the basis vectors for some of the sites are purely imaginary. The representation $\Gamma_{5}$ is two dimensional and repeated six times. It, therefore, corresponds to a magnetic structure of twelve basis vectors. The refinement of the magnetic structure was tested for each of these five IRs, and only $\Gamma_{5}$ gave satisfactory agreement. It may be noted that the two dimensionality of $\Gamma_{5}$ permits two physically equivalent descriptions of the magnetic structure. The basis vectors for $\Gamma_{5}$ indicate that the moment components along all $a$, $b$, and $c$ directions are refinable. The simultaneous refinement of all three moment components gives the values $m_a$ = 0.12 $\pm$ 0.08~$\mu_B$, $m_b$ = $-$0.09 $\pm$ 0.08~$\mu_B$, and $m_c$ = 2.25 $\pm$ 0.05~$\mu_B$. It is apparent that the moment components along the $a$ and $b$ axes are weak and in the order of the error bar. Therefore, the moments align predominantly along the $c$ axis which is the magnetic easy axis  [Fig. \ref{Fig:Mag}(a)] and the maximum possible deviation from the $c$ axis is less than 5 $\%$. The fitted pattern is shown in Fig. \ref{Fig:Neu-SCVO}(c) which shows good agreement with the observed pattern. Here a four phases model (nuclear and magnetic phases for both SrCo$_2$V$_2$O$_8$ and CoO) was used for the refinement. The $R_{mag}$-factor was found to be 5.08 $\%$ for the magnetic phase of SrCo$_2$V$_2$O$_8$. 

\begin{figure}
 \includegraphics[trim=0.8cm 9cm 0.5cm 0.5cm, clip=true, width=85mm]{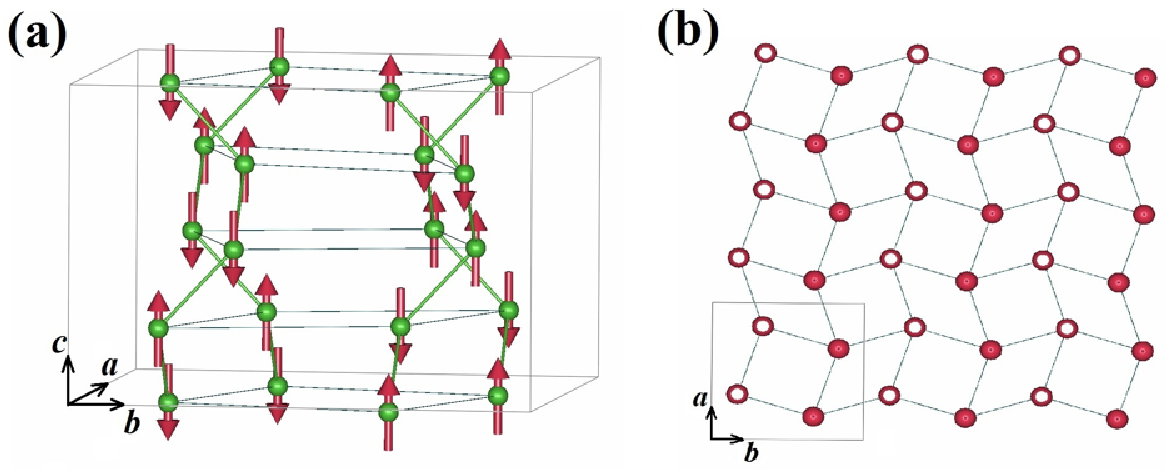}
 \caption{\label{Fig:Mag-struc-SCVO}(Color online) (a) The magnetic structure of SrCo$_2$V$_2$O$_8$. Thick solid green lines connect the spins within a chain along the {\it c} axis. The thin gray lines connect the spins within the {\it ab} plane. (b) The spin arrangements within a given {\it ab} plane (solid circle for spin up and open circle for spin down, respectively).}
 \end{figure}

The corresponding magnetic structure (weak moment components along the $a$ and $b$ axes are ignored) is shown in Fig.~\ref{Fig:Mag-struc-SCVO} where the Co-spins align antiferromagnetically along the chain ($c$ axis). Within the basal plane ($ab$ plane), the spins form ferromagnetic/antiferromagnetic zigzag lines along the $a/b$ axis. It may be noted that the observed magnetic structure is the same as that reported for the related compound BaCo$_2$V$_2$O$_8$.\cite{KawasakiPRB.83.064421} 

Further, to verify the magnetic structure, simulated annealing (SA) analysis \cite{CarvajalphysicaB.192.55} was performed using the FullProf suite program.\cite{Fullprof} The details of the SA method and the results are given in Appendix \ref{SA_SCVO-SMVO}. The solution of SA is the same as the  $\Gamma_{5}$ magnetic structure that was obtained from representation analysis with magnetic moments aligned along the $c$ axis [Table \ref{tab:T9}].

The refined value of the magnetic moment $\sim$ 2.25 $\mu_B$/Co$^{2+}$ at 1.5~K is substantially smaller than the value 3~$\mu_B$/Co$^{2+}$ expected for $S$ = 3/2 considering a spin only contribution. Therefore, it may be concluded that the Co moments are not fully ordered even at 1.5~K. This reduction is probably due to strong quantum fluctuations expected in this quasi-1D system. In fact, the presence of quantum fluctuations is evident from the field-induced transition from the 3D long-range AFM ordered state to the 1D quantum disordered state.\cite{HePRB.73.212406} Nevertheless, all magnetic Bragg peaks are found to be instrumental resolution limited, thus confirming that the AFM ordering is long-range at 1.5~K.

\subsubsection{Magnetic structure of SrMn$_2$V$_2$O$_8$}

\begin{figure} 
 \includegraphics[trim=4.2cm 4.7cm 13.5cm 0.5cm, clip=true, width=80mm]{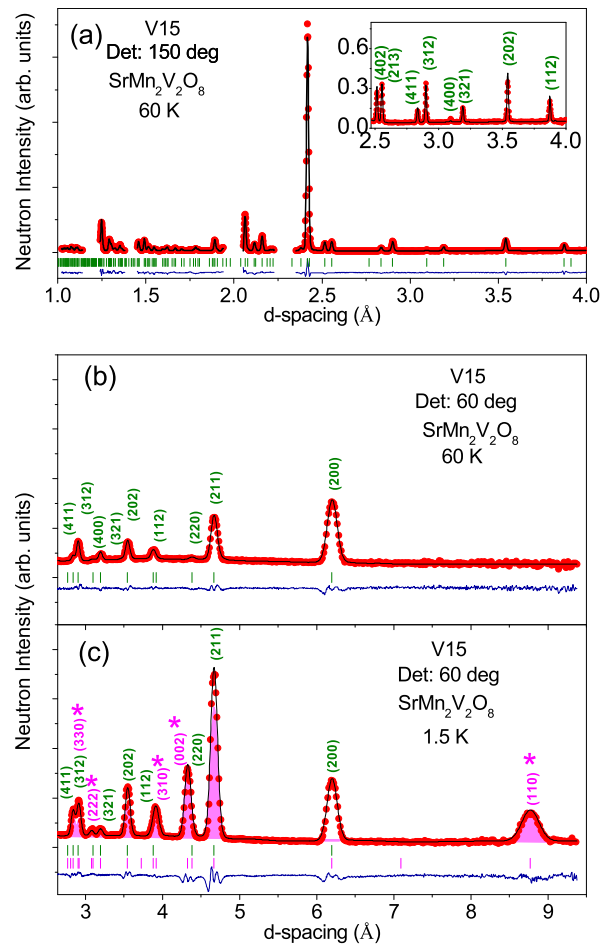}
 \caption{\label{Fig:Neu-SMVO}(Color online) The neutron diffraction patterns of SrMn$_2$V$_2$O$_8$ at (a) 60~K (from the detector bank at 150~deg), (b) 60~K (from the detector bank at 60~deg), and (c) 1.5~K (from the detector bank at 60~deg). The observed and calculated  patterns are shown by filled circles and solid black lines, respectively. The difference patterns between observed and calculated patterns are shown by solid lines at the bottom of each panel. The vertical bars are the calculated positions of the Bragg peaks. Different colors of the vertical bars correspond to the different phases. For the refinement of the diffraction pattern from the detector bank at 150~deg, shown in (a), few regions are excluded where the  aluminum lines (from the cryostat) were observed. The inset of (a) shows an enlarged view of the diffraction pattern over the high {\it d}-value region. The shaded regions in (c) show the contributions of the magnetic scattering from SrMn$_2$V$_2$O$_8$. The pure magnetic Bragg peaks are marked with stars.}
 \end{figure}

The neutron diffraction patterns of SrMn$_2$V$_2$O$_8$ at 60~K (paramagnetic state) and 1.5~K (magnetically ordered state) are shown in Fig. \ref{Fig:Neu-SMVO} (a)--(c). The diffraction pattern from the detector bank at 150~deg  [Fig. \ref{Fig:Neu-SMVO} (a)] at 60~K can be fitted by the nuclear phase alone with the tetragonal space group $I4_1cd$, as found in the room temperature diffraction study, confirming the paramagnetic state. The refinement shows good agreement between observed and calculated patterns. The refined values of the lattice parameter are $a$ = 12.3768 $\pm$ 0.0004~$\angstrom$ and $c$ = 8.6397 $\pm$  0.0003~$\angstrom$. The obtained values of the structural parameters were used to refine the diffraction patterns from the detector bank at 60~deg which covers high {\it d}-value (2.5--9.5~$\angstrom$) region and important for the magnetic scattering study. The refined diffraction pattern from this detector bank is depicted in Fig. \ref{Fig:Neu-SMVO}(b) which also shows good agreement between observed and calculated patterns. 

The diffraction pattern measured at 1.5 K (in the magnetic ordered state) is shown in Fig. \ref{Fig:Neu-SMVO}(c). Appearance of a set of additional magnetic Bragg peaks as well as the increase of the intensity of few nuclear Bragg peaks have been observed. Unlike SrCo$_2$V$_2$O$_8$, the magnetic peaks for this compound are at positions ({\it hkl}) where $h+k+l$ is an even integer, that is, on the nuclear peaks allowed by the body-centered symmetry {\it I} of the lattice, however, forbidden by the space group symmetry $I4_1cd$. Therefore, the propagation vector {\it k} = (0 0 0) can be used for the magnetic phase. 

For this case, we have also carried out representation analysis to find out the symmetry allowed magnetic structures. As discussed earlier in the case of SrCo$_2$V$_2$O$_8$, the space group $I4_1 c d$ involves two centering operations and eight symmetry operations (Appendix \ref{G0_elements}). All the eight symmetry operations leave the propagation vector $k$ invariant. For the propagation vector $k$ = (0 0 0), the IRs of the propagation vector group $G_k$ are given in Table \ref{tab:T5}. The decomposition of the magnetic representation $\Gamma_{mag}$ in terms of the non-zero IRs of $G_k$ for the Mn site (16$b$ site) is
 
\begin{equation}
\Gamma_{mag}=3\Gamma_{1}^{1}+3\Gamma_{2}^{1}+3\Gamma_{3}^{1}+3\Gamma_{4}^{1}+6\Gamma_{5}^{2}
\label{magnetic_representation_SMVO}
\end{equation}
\noindent The associated basis vectors for the Mn site [16$b$ ($x$, $y$, $z$); (0.3328, 0.3184, 0.2226)] for all five IRs are given in Table \ref{tab:T6}.

\begin{figure} [bp]
 \includegraphics[trim=0.5cm 11cm 0.5cm 0cm, clip=true, width=85mm]{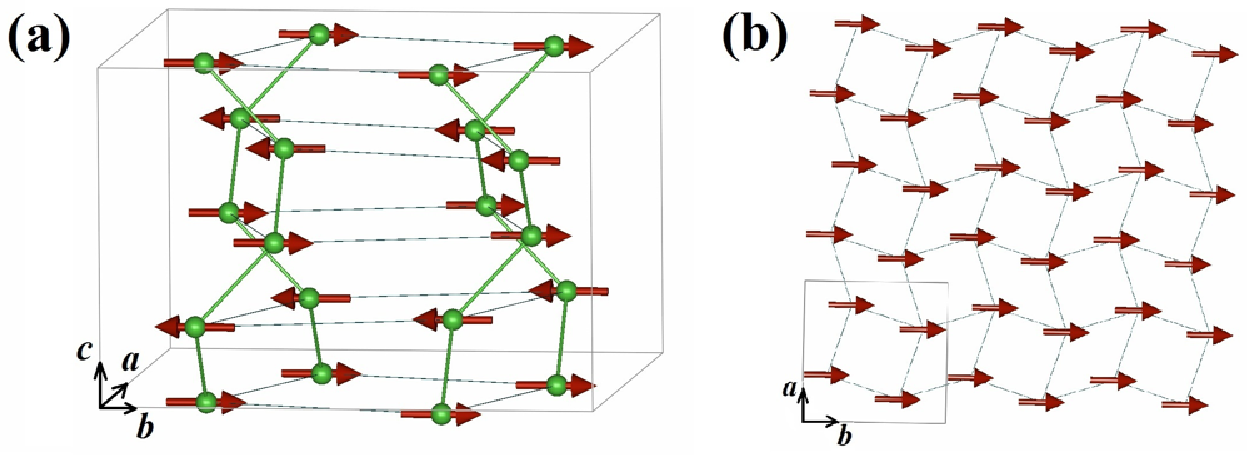}
 \caption{\label{Fig:Mag-struc-SMVO}(Color online) (a)  The schematic magnetic structure (considering the moments along the $b$ axis) of SrMn$_2$V$_2$O$_8$. Thick solid green lines connect the spins within a chain along the {\it c} axis. The thin gray lines connect the spins within a given {\it ab} plane. (b) The ferromagnetic spin arrangements within a given {\it ab} plane.}
 \end{figure}

The representations $\Gamma_{1}$, $\Gamma_{2}$, $\Gamma_{3}$, and $\Gamma_{4}$ are one dimensional and repeated three times. The representation $\Gamma_{5}$ is two dimensional and repeated six times. The refinement of the magnetic structure was tested for each of the irreducible representations, and only $\Gamma_{5}$ gave satisfactory agreement. The basis vectors for $\Gamma_{5}$ indicate that the moment components along all $a$, $b$, and $c$ directions are refinable. The refinement confirms that the moments are aligned in the $ab$ plane with $m_c$ component being zero or very weak (should be less than 5 $\%$ if any).  Calculated pattern considering any arbitrary moment direction within the $ab$ plane fits the observed data ($R_{mag}$ = $7.3 - 7.38 ~\%$). However, the refinement with the moment along the $a$ or $b$ axis and the other components fixed to zero gives a marginally better fit ($R_{mag}$ = $7.3 \%$). Therefore, it is not possible to determine the moment direction uniquely within the $ab$ plane from the present powder data. Single crystal neutron diffraction is required to determine the moment direction. Nevertheless, this powder diffraction investigation confirms that the moments are arranged parallel to each other within the $ab$ plane and antiparallel to each other along the $c$ axis. The calculated pattern considering moments entirely along the $a$ or $b$ axis is shown along with observed pattern in Fig. \ref{Fig:Neu-SMVO}(c).  A schematic magnetic structure, considering the moments along the $b$ axis, is shown in Fig.~\ref{Fig:Mag-struc-SMVO}.

\begin{table}
\caption{\label{tab:T5}Irreducible representations of the group $G_k$ of the propagation vector $k$ = (0 0 0) for SrMn$_2$V$_2$O$_8$. The atoms of the nonprimitive basis are defined according to Mn1: (0.3328, 0.3184, 0.2226), Mn2: (0.6672, 0.6816, 0.2226), Mn3: (0.6816, 0.8328, 0.4726), Mn4: (0.3184, 0.1672, 0.4726), Mn5: (0.6672, 0.3184, 0.7226), Mn6: ( 0.3328, 0.6816, 0.7226), Mn7: (0.8184, 0.3328, 0.4726), Mn8: (0.1816, 0.6672, 0.4726).}
\begin{ruledtabular}
\begin{tabular}{ccccccccc}
IRs & & & & Atoms & & & & \\
& Mn1 & Mn2 & Mn3 & Mn4 & Mn5 & Mn6 & Mn7 & Mn8 \\
\hline
$\Gamma_1^1$ & 1 & 1 & 1 & 1 & 1 & 1 & 1 & 1 \\ \\
$\Gamma_2^1$ & 1 & 1 & 1 & 1 &  -1 & -1 & -1 & -1 \\ \\
$\Gamma_3^1$ & 1 & 1 & -1 & -1 & 1 & 1 & -1 & -1 \\ \\
$\Gamma_4^1$ & 1 & 1 &  -1 & -1 &  -1 &  -1 & 1 & 1 \\ \\
$\Gamma_5^2$ & 1  0 & -1  0 & 0  1 & 0  -1 & 1  0 &-1  0 & 0  1 & 0  -1 \\
 & 0 1 & 0  -1 &  -1  0 &1  0  &  0  -1 & 0  1 &  1  0 & -1  0 \\
\end{tabular}
\end{ruledtabular}
\end{table}

\begin{table}
\caption{\label{tab:T6}Basis Vectors of position 16$b$ for the IRs with $k$ = (0 0 0) for SrMn$_2$V$_2$O$_8$. Only the real components of the basis vectors are presented. The atomic sites are labeled following the convention given in Table \ref{tab:T5}.}
\begin{ruledtabular}
\begin{tabular}{cccccccccc}
IRs &&\multicolumn{8}{c}{Basis Vectors} \\
& & Mn1 & Mn2 & Mn3 & Mn4 & Mn5 & Mn6 & Mn7 & Mn8 \\
\hline
$\Gamma_1^1$ & $\Psi_1$ &(100)&(-100)&(010)&(0-10)&(100)	&(-100)&(0-10)&	(010)\\
 & $\Psi_2$ &(010)&	(0-10)&	(-100)&	(100)&	(0-10)&	(010)&	(-100)&	(100)\\
 & $\Psi_3$ &(001)&	(001)&	(001)&	(001)&	(00-1)&	(00-1)&	(00-1)&	(00-1)\\
\\
$\Gamma_2^1$ &$\Psi_1$ &(100)&	(-100)&	(010)&	(0-10)&	(-100)&	(100)&	(010)&	(0-10)\\
&$\Psi_2$&(010) &	(0-10) &	(-100) &	(100) &	(010) &	(0-10) &	(100) &	(-100)\\
&$\Psi_3$&(001)&	(001)&	(001)&	(001)&	(001)&	(001)&	(001)&	(001) \\
\\
$\Gamma_3^1$ & $\Psi_1$ & (100)&	(-100)&	(0-10)&	(010)&	(100)&	(-100)&	(010)&	(0-10) \\
&$\Psi_2$&(010)&	(0-10)&	(100)&	(-100)&	(0-10)&	(010)&	(100)&	(-100) \\
&$\Psi_3$&(001)&	(001)&	(00-1)&	(00-1)&	(00-1)&	(00-1)&	(001)&	(001) \\
\\
$\Gamma_4^1$ &  $\Psi_1$  & (100)&	(-100)&	(0-10)&(010)&	(-100)&	(100)&	(0-10)&	(010) \\
& $\Psi_2$&(010)&	(0-10)&	(100)&	(-100)&	(010)&	(0-10)&	(-100)&	(100) \\
& $\Psi_3$&(001)&	(001)&	(00-1)&	(00-1)&	(001)&	(001)&	(00-1)&	(00-1) \\
\\
$\Gamma_5^2$ &$\Psi_1$& (100)&	(100)&	(000)&	(000)&	(100)&	(100)&	(000)&	(000)\\
&$\Psi_2$&(010)&	(010)&	(000)&	(000)&	(0-10)&	(0-10)&	(000)&	(000) \\
&$\Psi_3$&(001)&	(00-1)&	(000)&	(000)&	(00-1)&	(001)&	(000)&	(000) \\
&$\Psi_4$&(000)&	(000)&	(010)&	(010)&	(000)&	(000)&	(0-10)&	(0-10) \\
&$\Psi_5$&(000)&	(000)&	(-100)&	(-100)&	(000)&	(000)&	(-100)&	(-100) \\
&$\Psi_6$&(000)&	(000)&	(001)&	(00-1)&	(000)&	(000)&	(00-1)&	(001) \\
&$\Psi_7$&(000)&	(000)&	(0-10)&	(0-10)&	(000)&	(000)&	(0-10)&	(0-10) \\
&$\Psi_8$&(000)&	(000)&	(100)&	(100)&	(000)&	(000)&	(-100)&	(-100) \\
&$\Psi_9$&(000)&	(000)&	(00-1)&	(001)&	(000)&	(000)&	(00-1)&	(001) \\
&$\Psi_{10}$&(100)&	(100)&	(000)&	(000)&	(-100)&	(-100)&	(000)&	(000) \\
&$\Psi_{11}$&(010)&	(010)&	(000)&	(000)&	(010)&	(010)&	(000)&	(000) \\
&$\Psi_{12}$&(001)&	(00-1)&	(000)&	(000)&	(001)&	(00-1)&	(000)&	(000) \\
\end{tabular}
\end{ruledtabular}
\end{table}

For further verification of the magnetic structure, we have performed an SA analysis (see Appendix \ref{SA_SCVO-SMVO}). The solution of the SA gives a magnetic structure where spins are preferably arranged antiferromagnetically along the $c$ axis and ferromagnetically within a given $ab$ plane. The solution also indicates that the moments are preferably aligned perpendicular to the $c$ axis and mostly along the $a$ or $b$ axis. The solution of SA, therefore, agrees well with the magnetic structure obtained from the representation analysis (corresponds to the $\Gamma_{5}$). 

The refined value of the magnetic moment is found to be 3.99 $\pm$ 0.01~$\mu_B$/Mn$^{2+}$ at 1.5~K. This value is quite small compared to the fully ordered moment of 5~$\mu_B$ (Mn$^{2+}$: 3$d^5$; $S = 5/2$). The reduced ordered moment even at 1.5~K (T/T$_N$ $\approx$ 0.036) suggests the presence of quantum fluctuations due to the quasi-1D magnetic interactions. However, all magnetic peaks are found to be instrumental resolution limited suggesting 3D long-range AFM order at 1.5~K.

Both the magnetization and neutron diffraction studies reveal that the two iso-structural compounds   SrCo$_2$V$_2$O$_8$ and SrMn$_2$V$_2$O$_8$ have different magnetic ground states. Unlike SrMn$_2$V$_2$O$_8$, the magnetic structure of SrCo$_2$V$_2$O$_8$, which consists of a  ferro-/antiferromagnetic arrangement of AFM chains within the tetragonal plane ($ab$ plane) is unusual for a tetragonal structure. Two explanations for this situation may be considered as (i) the presence of competing interchain interactions along the diagonal ($\langle110\rangle$) direction within the $ab$ plane and/or (ii) a structural transition from the tetragonal to an orthorhombic symmetry which accompanies the magnetic ordering. 

For the former case, the competition between AFM $J'_{NN}$ (along the principle axes $a$ and $b$) and AFM $J'_{NNN}$ (along the diagonal direction in the $ab$ plane) results in magnetic frustration for a  square-lattice (tetragonal symmetry). For a frustrated square-lattice system, the $J_1$-$J_2$ model (where $J_1$ and $J_2$ are the exchange interactions along the side and diagonal of the square), predicts a collinear stripe-like AFM structure for $J_2$/$J_1$ $\ge$ 0.6. The collinear stripe structure is occurred by an order by disorder phenomenon.\cite{SchmalfussPRL.97.157201} However, it should be noted that the spin-geometry within a given $ab$ plane for the studied compounds is more complex [Fig. \ref{Fig:CrysStruc}(c)] than a simple square-lattice. Furthermore, the dominant interaction is the intrachain AFM interaction which is expected to be an order of magnitude greater than the interchain interactions. Moreover, the crystal structure suggests the possibility of a complex interchain interaction topology with many competing interactions; not only within the $ab$ plane but also with an out-of-plane component. The screw chains also may allow second-nearest-neighbor intrachain interactions along the $c$ axis. To investigate the interchain interactions in detail, inelastic neutron scattering studies on single crystal sample are required. 

The reduction of the magnetic symmetry can also be explained by a finite magnetoelastic coupling within the $ab$ plane. In this case, the effective $J'_{NN}$ interactions along two principle axes become unequal due to lattice distortions. However, the lattice distortions, if any, are rather weak and, therefore, not evident in the present neutron diffraction data which has moderate resolution. Note that, neither the sign nor the magnitude of the change of $J'_{NN}$ as a function of distance are known. To search for a possible distortion, high resolution diffraction studies as a function of temperature by using either neutron or synchrotron radiation are desired.

\subsection{Critical exponent}

\begin{figure}
\includegraphics[trim=2.5cm 4cm 13cm 1.5cm, clip=true, width=70mm]{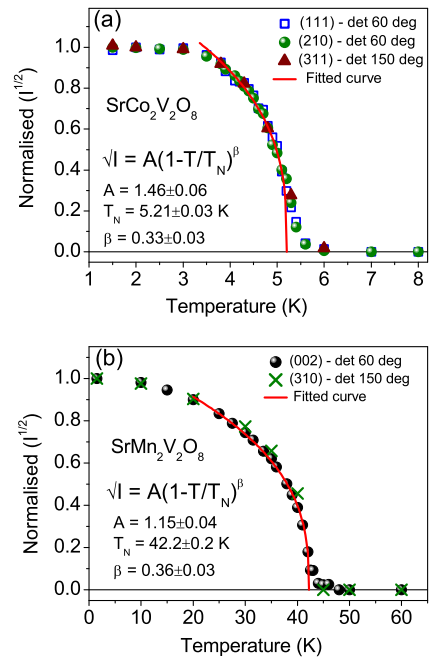}
\caption{\label{Fig:criti-expt}(Color online) The temperature dependence of the normalized square root of integrated intensity of (a) the (111), (210), and (311) magnetic Bragg peaks for SrCo$_2$V$_2$O$_8$, and (b) the (002) and (310) magnetic Bragg peaks for SrMn$_2$V$_2$O$_8$. The solid red lines are the fitted curves according to Eq. \ref{eq:criti} as described in the text.}
 \end{figure}
In order to understand further the nature of magnetic orderings, we have carried out a temperature-dependent neutron diffraction study. Figures \ref{Fig:criti-expt} (a) and (b) show the temperature dependence of the normalized square root of the integrated intensities of a few selected magnetic Bragg peaks of SrCo$_2$V$_2$O$_8$ and SrMn$_2$V$_2$O$_8$, respectively. The integrated intensities were obtained from Gaussian function fit of the magnetic satellite peaks. The fits confirm that the positions of the magnetic peaks do not change with temperature proving that there is no change of the $k$-vectors for both compounds. The intensity of each magnetic peak was normalized with respect to its intensity at the base temperature 1.5~K. These measurements allowed us to determine the critical exponent $\beta$ for the temperature induced phase transition from the long-range AFM ordered state to the paramagnetic state. 

The order parameter (i.e., the ordered AFM moment {\it m}) can be expressed as $m(T) \propto (T_N - T)^\beta$. The ordered magnetic moment is proportional to the square root of the integrated intensity of the magnetic peaks. Thus the critical exponent $\beta$ can be obtained by fitting the square root of the integrated intensity to the following equation 
\begin{eqnarray}
\sqrt{I(T)} = A(1-T/T_N)^\beta
\label{eq:criti}
\end{eqnarray} 
\noindent where {\it A} is a proportionality constant and $T_N$ is the 3D AFM ordering temperature. For SrCo$_2$V$_2$O$_8$, the fit of the above equation to the observed data over a limited temperature range near $T_N$ ($0.75 \le T/T_N \le 1$) is shown by solid red curve in Fig. \ref{Fig:criti-expt}(a). The fitted values of $\beta$ and $T_N$ are found to be 0.33 $\pm$ 0.03 and 5.21 $\pm$ 0.03~K, respectively. The fitted value of $\beta$ = 0.33 is close to the value $\beta \approx 5/16$ $(= 0.3125)$ predicted for the 3D Ising model. \cite{Blundell.book} The slightly higher value of $\beta$ may be due to the deviation from the pure Ising limit as evident in our temperature-dependent susceptibility study. For SrMn$_2$V$_2$O$_8$, the critical exponent $\beta$ and $T_N$  are 0.36 $\pm$ 0.03 and 42.2 $\pm$ 0.02~K, respectively, obtained from the fit of the data in the temperature range $0.5 \le T/T_N \le 1$  [Fig. \ref{Fig:criti-expt}(b)]. The fitted value of $\beta$ agrees well with the theoretically expected value of 0.367 for the 3D Heisenberg model.\cite{Blundell.book}

\subsection{Short-range spin-spin correlations}

\begin{figure}
\includegraphics[trim=2cm 5.5cm 5cm 1.5cm, clip=true, width=85mm]{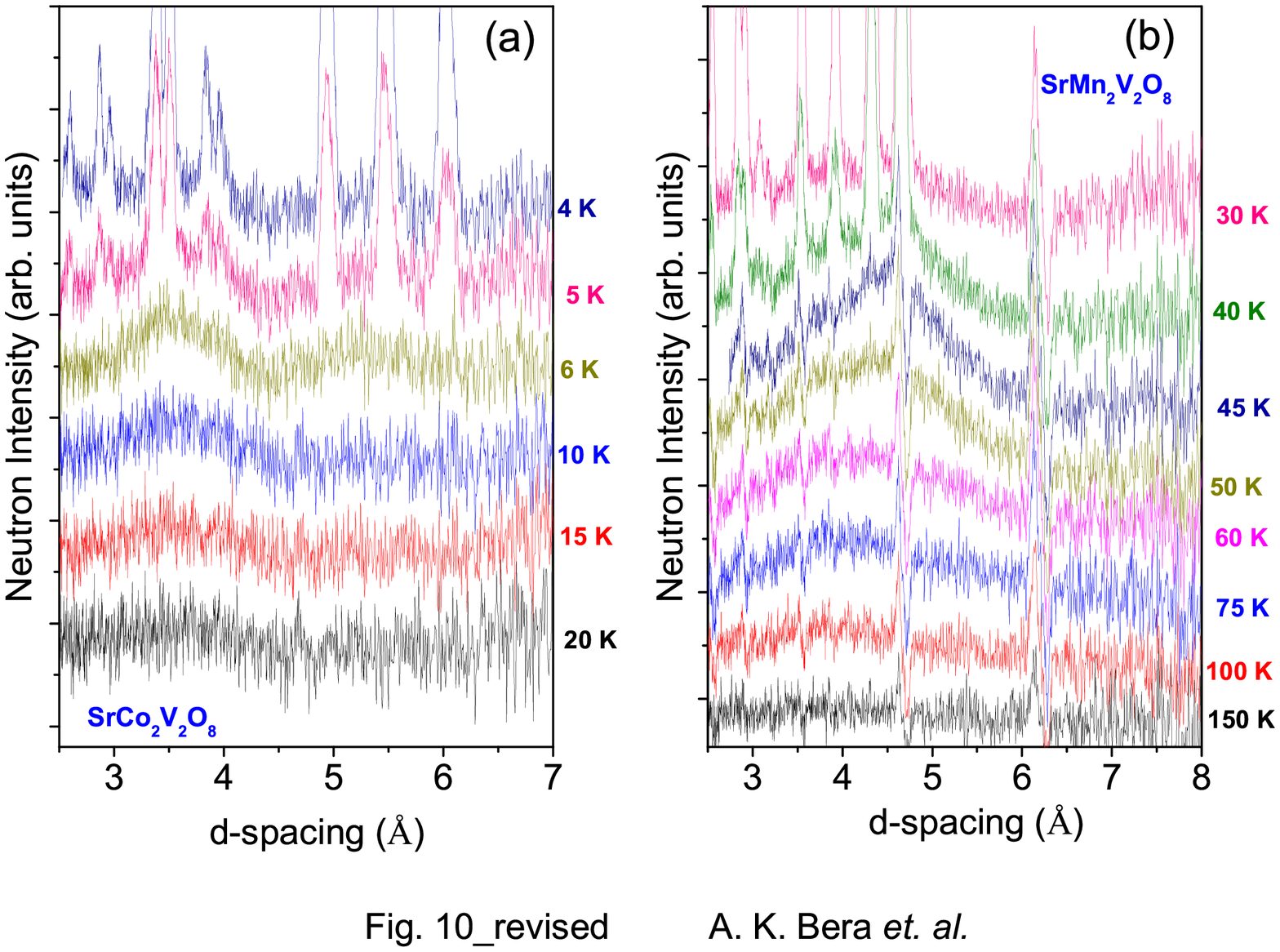}
\caption{\label{Fig:Diffscat}(Color online) Magnetic scattering for (a) SrCo$_2$V$_2$O$_8$ and (b) SrMn$_2$V$_2$O$_8$, after subtraction of the nuclear background at 30~K and 200~K, respectively. The patterns are shifted vertically for the purpose of the presentation. Broad diffuse magnetic scattering, corresponding to short-range spin-spin correlations, are evident for both compounds over a wide temperature range.}
 \end{figure}

Now, we focus on the nature of the magnetic correlations above and around $T_N$. The magnetic patterns at 20, 15, 10, 6, 5, and 4~K for SrCo$_2$V$_2$O$_8$ and at 150, 100, 75, 60, 50, 45, 40, and 30~K for SrMn$_2$V$_2$O$_8$ are shown in Figs. \ref{Fig:Diffscat} (a) and (b), respectively. 
The magnetic patterns were obtained by subtraction of the paramagnetic background at 30~K for the Co- and at 200~K for the Mn-compound, respectively. For both compounds, the presence of diffuse magnetic scattering due to short-range spin-spin correlations is evident over a wide temperature range above and below $T_N$. This is in agreement with the $\chi$(T) study where broad peaks, indicating the short-range spin-spin correlations, appeared centered at $\sim$ 30~K for the Co-compound and at $\sim$ 170~K for the Mn-compound, respectively [Fig. \ref{Fig:Sus}]. 

For the Co-compound, below 20~K, a broad diffuse peak appears at $d \sim 3.4$~$\angstrom$. On lowering the temperature, the peak intensity increases continuously. At $T \le$ 6~K, an additional broad peak appears at $d \sim 5.2$~$\angstrom$. When the temperature is lowered further, the broad peaks gradually transform into a set of Bragg peaks below $T_N \sim 5.2$~K. The complete transformation occurs at $\sim$ 4~K resulting in a coexistence of both broad diffuse peaks and sharp Bragg peaks over the temperature range 5.2--4~K. For the Mn-compound, with decreasing temperature below 150~K a broad diffuse peak appears at $d \sim 4.7$~$\angstrom$. With further lowering of temperature, the peak intensity increases continuously and gradually transforms into Bragg peaks below $T_N \sim 42.2$~K. The  broad diffuse peak is found to coexist below $T_N$ down to $\sim$ 30~K.

From this diffuse scattering study, it is evident that the transition from short-range to long-range magnetic ordering is not sharp but rather occurs gradually over a wide temperature range in both compounds. The short-range spin-spin correlations are expected to be one dimensional (due to the strong intrachain interactions) and appear when the interaction strength becomes comparable to the thermal energy. With further lowering of temperature, the thermal energy becomes comparable to the interchain interaction strength and 3D long-range magnetic ordering starts to occur at $T_N$. To investigate the exact nature of the short-range correlations, neutron diffraction studies on single crystal are required. Nevertheless, this study confirms the presence of short-range correlations over a wide temperature range above and below $T_N$ for both compounds.

\section{Summary and conclusion}
In summary, we have investigated magnetic correlations in two iso-structural compounds Sr$M_2$V$_2$O$_8$ with different magnetic ions Co$^{2+}$ and Mn$^{2+}$ by using dc-magnetization and neutron diffraction. The crystal structures of both compounds have tetragonal symmetry with space group $I4_1cd$. However, an expansion of unit cell volume from the Co-compound to the Mn-compound has been found due to the ionic radii effect. Both magnetization and neutron diffraction studies reveal that two compounds have different magnetic properties and different AFM ground states. For the Mn-compound, AFM chains (along the {\it c} axis) are ordered ferromagnetically within the {\it ab} plane, whereas, for the Co-compound, AFM chains are ordered ferro-/antiferromagnetically along  the {\it a/b} direction. Reduced ordered moments have been found for both compounds at base temperature 1.5~K revealing the presence of strong quantum fluctuations due to the quasi-1D magnetic interactions. Critical exponent study indicates that the Co- and Mn-compounds belong to the Ising and Heisenberg universality classes, respectively. For both compounds, the presence of short-range spin-spin correlations have been found over a wide temperature range due to the quasi-one-dimensional magnetic interactions.


\appendix

\section{Elements in the group G$_0$}
\label{G0_elements}
The symmetry elements of the space group $I 4_1 c d$ (group G$_0$) is given in Table \ref{symmetry elements_of_G0_table}.

\begin{table*}
\caption{\label{symmetry elements_of_G0_table} Symmetry operators of the space group $I 4_1 c d$. The notations used are of the International Tables, where the elements are separated into rotation\cite{Bradley-Cracknell.book} and translation components, and the {\it Jones faithful representations}.\footnote{While the {\it Jones faithful representations} conventionally correspond to the vector formed from the operation of the rotation part of the element, {\it R}, on the site coordinates $(x,~y,~z)$, here we use it to demonstrate the effect of the complete symmetry operator, {\it g}.} All the eight symmetry operations leave the propagation $k$ invariant or transform it into an equivalent vector.}

\begin{ruledtabular}
\begin{tabular}{ccccc}
Element & Rotation matrix & IT notation  & Kovalev notation & Jones  symbol \\ 
 $g_n$     & {\it R}    & $g_n=\{R\mid \bftau$ \}& $g_n=\{h_n\mid \bftau$ \} &           \\ 
\hline
$g_{1}$ & $ \left( \begin{array}{rrr}  1 &  0 &  0  \\  0 &  1 &  0  \\  0 &  0 &  1  \end{array}  \right) $   & 
$\{E |~0~0~0\}$ & $\{h_{1} \mid ~0~0~0\}$ & $x, y, z$ \\
$g_{2}$ & $ \left( \begin{array}{rrr}  \bar{1} &  0 &  0  \\  0 &  \bar{1} &  0  \\  0 &  0 &  1  \end{array}  \right) $   & 
$\{C_{2z} |~0.5~0.5~0.5\}$ & $\{h_{4} \mid ~0.5~0.5~0.5\}$ & $-x+\frac{1}{2}, -y+\frac{1}{2}, z+\frac{1}{2}$ \\
$g_{3}$ & $ \left( \begin{array}{rrr}  0 &  \bar{1} &  0  \\  1 &  0 &  0  \\  0 &  0 &  1  \end{array}  \right) $   & 
$\{C_{4z}^{+} |~0~0.5~0.25\}$ & $\{h_{14} \mid ~0~0.5~0.25\}$ & $-y, x+\frac{1}{2}, z+\frac{1}{4}$ \\
$g_{4}$ & $ \left( \begin{array}{rrr}  0 &  1 &  0  \\  \bar{1} &  0 &  0  \\  0 &  0 &  1  \end{array}  \right) $   & 
$\{C_{4z}^{-} |~0.5~0~0.75\}$ & $\{h_{15} \mid ~0.5~0~0.75\}$ & $y+\frac{1}{2}, -x, z+\frac{3}{4}$ \\
$g_{5}$ & $ \left( \begin{array}{rrr}  1 &  0 &  0  \\  0 &  \bar{1} &  0  \\  0 &  0 &  1  \end{array}  \right) $   & 
$\{\sigma_{y} |~0~0~0.5\}$ & $\{h_{27} \mid ~0~0~0.5\}$ & $x, -y, z+\frac{1}{2}$ \\
$g_{6}$ & $ \left( \begin{array}{rrr}  \bar{1} &  0 &  0  \\  0 &  1 &  0  \\  0 &  0 &  1  \end{array}  \right) $   & 
$\{\sigma_{x} |~0.5~0.5~0\}$ & $\{h_{26} \mid ~0.5~0.5~0\}$ & $-x+\frac{1}{2}, y+\frac{1}{2}, z$ \\
$g_{7}$ & $ \left( \begin{array}{rrr}  0 &  \bar{1} &  0  \\  \bar{1} &  0 &  0  \\  0 &  0 &  1  \end{array}  \right) $   & 
$\{\sigma_{da} |~0~0.5~0.75\}$ & $\{h_{40} \mid ~0~0.5~0.75\}$ & $-y, -x+\frac{1}{2}, z+\frac{3}{4}$ \\
$g_{8}$ & $ \left( \begin{array}{rrr}  0 &  1 &  0  \\  1 &  0 &  0  \\  0 &  0 &  1  \end{array}  \right) $   & 
$\{\sigma_{db} |~0.5~0~0.25\}$ & $\{h_{37} \mid ~0.5~0~0.25\}$ & $y+\frac{1}{2}, x, z+\frac{1}{4}$ \\
\end{tabular}
\end{ruledtabular}
\end{table*}

\section{Simulated Annealing for SrCo$_2$V$_2$O$_8$ and SrMn$_2$V$_2$O$_8$}
\label{SA_SCVO-SMVO}

\begin{table}
\caption{\label{tab:T9}Crystallographic components of Fourier Coefficients of Magnetic Moments as obtained from SA analysis for SrCo$_2$V$_2$O$_8$. The atomic sites are labeled following the convention given in Table \ref{tab:T3}.}
\begin{ruledtabular}
\begin{tabular}{ccccccc}
Atoms &$R$ &$\phi$ &$\theta$ & $R_x$ & $R_y$&$R_z$ \\
\hline
Co1 & 2.0826&	314.0477&   	0.0000& 	0.0000&    	0.0000 &   	2.0826 \\
Co2 & 2.0826&	233.6902& 	180.0000&	0.0000&    	0.0000&    	-2.0826 \\
Co3 & 2.0826&	228.9207&   	0.0001&	0.0000&    	0.0000&    	2.0826 \\
Co4 & 2.0826&	152.6009&	170.2723&	-0.3124&   	0.1619&    	-2.051 \\
Co5 & 2.0826&	335.5125&   	0.0000&	0.0000& 	0.0000&	2.0826 \\
Co6 & 2.0826&	168.1992& 	180.0000&	0.0000&    	0.0000&   	-2.0826 \\
Co7 & 2.0826&	317.1950& 	180.0000&	0.0000&    	0.0000&   	-2.0826 \\
Co8 & 2.0826&	147.9146&	9.1028&	 -0.2792&    	 0.1750&    	2.056\\
$R_F^2$ & 7.1 $\%$ &&&&&\\
\end{tabular}
\end{ruledtabular}
\end{table}

\begin{table}
\caption{\label{tab:T10}Crystallographic components of Fourier Coefficients of Magnetic Moments as obtained from SA analysis for SrMn$_2$V$_2$O$_8$. The atomic sites are labeled following the convention given in Table \ref{tab:T5}.}
\begin{ruledtabular}
\begin{tabular}{ccccccc}
Atoms &$R$ &$\phi$ &$\theta$ & $R_x$ & $R_y$&$R_z$ \\
\hline
Mn1 & 3.8724&	99.6550&  	88.2367&	 -0.649&   	  3.816&   	0.119\\
Mn2 & 3.8724&	102.4641&  	98.3593&	-0.827& 	3.741&   	-0.563 \\
Mn3 & 3.8724&	259.4255&  	82.7192&	-0.705&    	-3.776&   	0.491\\
Mn4 & 3.8724&	275.7251&	101.6911&	0.378&    	-3.773&   	-0.785 \\
Mn5 & 3.8724&	97.9587&  	95.4424& 	-0.534&  	3.818&    	-0.367 \\
Mn6 & 3.8724&	81.7089&  	94.9278&	0.556&  	3.818&   	-0.333\\
Mn7 & 3.8724&	275.1613&  	89.0647&	0.3483&    	-3.856&    	0.063 \\
Mn8 & 3.8724&	262.0538&	88.2306&	-0.535&  	-3.833&    	0.119\\
$R_F^2$ & 10.1 $\%$ &&&&&\\
\end{tabular}
\end{ruledtabular}
\end{table}

The simulated annealing (SA) analysis was performed using the FullProf program suite\cite{Fullprof} to verify the magnetic structures obtained from the representation analyses. This is a very general method to determine the magnetic structure. It uses an adaption of the Metropolis algorithm (a Monte Carlo method) to generate sample states of a thermodynamic system. It does not depend on the initial parameters and is therefore ideal for solving complex and/or completely unknown structures. In detail, the SA method is based on the concept of minimizing a $cost$ function $E$($\omega$) with respect to a given configuration vector $\omega$.\cite{CarvajalphysicaB.192.55,KirkpatrickScience.220.671} 

For the input of the SA analysis, the integrated intensities of the magnetic peaks were determined from the 1.5~K patterns by using the mixed Rietveld and profile matching methods in the Fullprof program. Here, all structural parameters for the nuclear phases were fixed (please note that for SrCo$_2$V$_2$O$_8$,  additional  parameters for nuclear and magnetic phases of CoO were also fixed) and the magnetic contributions were introduced as a new phase using the Le Bail fit mode. The SA analysis was, then, done without considering any symmetry constrain. The only constrain was the same magnitude of the moments for all eight independent magnetic sites within a unit cell. This gives total 17 refinable parameters; one magnetic moment parameter and 8$\times$2 = 16 angular parameters. One of the solutions each for  SrCo$_2$V$_2$O$_8$ and SrMn$_2$V$_2$O$_8$ are listed in Table \ref{tab:T9} and Table \ref{tab:T10}, respectively. 

The solution for SrCo$_2$V$_2$O$_8$ corresponds to a magnetic structure in which the moments are arranged antiferromagntically along the $c$ axis and ferro/antiferro-magnetically along the $a/b$ axis with moments aligned along the $c$ axis.  

The solution for SrMn$_2$V$_2$O$_8$ corresponds to a magnetic structure where moments make an almost parallel arrangement within a given $ab$ plane and an antiparallel arrangement of such planes along the $c$ axis. The moments are aligned mostly along the $a/b$ axis.

\newpage


%

\end{document}